\documentclass[conference]{IEEEtran}
\usepackage{fancybox}
\usepackage[ruled,vlined]{algorithm2e}
\usepackage{graphicx}
\usepackage{paralist}
\usepackage{tabularx}
\usepackage[hyphens]{url}
\usepackage{balance}
\usepackage{multirow}
\usepackage{multicol}
\usepackage{pbox}
\usepackage{mathtools}
\usepackage{cite}
\usepackage{adjustbox}
\usepackage[TABBOTCAP]{subfigure}
\usepackage{url}
\usepackage{balance}
\usepackage{amsmath}
\usepackage{verbatim}
\usepackage{float}
\usepackage{colortbl}
\usepackage{color}
\usepackage{amssymb}
\usepackage{ifthen}
\usepackage{pifont}
\usepackage{tikz}
\newcommand*\circled[1]{\tikz[baseline=(char.base)]{
            \node[shape=circle,draw,inner sep=0.5pt] (char) {#1};}}

\newcommand{\CrashScopesp}{{\sc CrashScope~}}
\newcommand{\CrashScope}{{\sc CrashScope}}
\newcommand{\CrashScopes}{{\sc CrashScope's~}}
\newcommand{\CrashDroid}{{\sc CrashDroid~}}

\include{macro}

\begin{document}

\title{Automatically Discovering, Reporting and Reproducing Android Application Crashes}

\author{\IEEEauthorblockN{Kevin Moran, Mario Linares-V\'asquez, Carlos Bernal-C\'ardenas, Christopher Vendome, and Denys Poshyvanyk}
\IEEEauthorblockA{College of William \& Mary\\
\{kpmoran, mlinarev, cebernal, cvendome, denys\}@cs.wm.edu}}

\maketitle
\vspace{-0.5cm}
\begin{abstract}
Mobile developers face unique challenges when detecting and reporting crashes in apps due to their prevailing GUI event-driven nature and additional sources of inputs (e.g., sensor readings).  To support developers in these tasks, we introduce a novel, automated approach called \CrashScope.  This tool explores a given Android app using systematic input generation, according to several strategies informed by static and dynamic analyses, with the intrinsic goal of triggering crashes. When a crash is detected, \CrashScopesp generates an augmented crash report containing screenshots, detailed crash reproduction steps, the captured exception stack trace, and a fully replayable script that automatically reproduces the crash on a target device(s). 

We evaluated \CrashScope's effectiveness in discovering crashes as compared to five state-of-the-art Android input generation tools on 61 applications. The results demonstrate that \CrashScopesp performs about as well as current tools for detecting crashes and provides more detailed fault information. Additionally, in a study analyzing eight real-world Android app crashes, we found that  \CrashScope's reports are easily readable and allow for reliable reproduction of crashes by presenting more explicit information than human written reports.\end{abstract}

\vspace{-0.2cm}
\section{Introduction}
\label{sec:intro}
\vspace{-0.15cm}
Continued growth in the mobile hardware and application marketplace is being driven by a landscape where users tend to prefer mobile smart devices and apps for tasks over their desktop counterparts. 
The gesture-driven nature of mobile apps has given rise to new challenges encountered by programmers during development and maintenance, specifically with regard to testing and debugging \cite{23Joorabchi:ESE13}.   One of the most difficult \cite{3Bettenburg:FSE08,15Breu:CSCW10} and important maintenance tasks is the creation and resolution of bug reports\cite{36Gu:ICSE10}.  
Reports concerning application crashes are of particular importance to developers, because crashes represent a jarring software fault that is directly user facing and immediately impacts an app's utility and sucess.   If an app is not behaving as expected due to crashes, missing features, or other bugs, nearly half of users are likely to abandon the app for a competitor \cite{app-abandonment} in a marketplace such as Google Play \cite{google-play}.

Mobile developers heavily rely on user reviews \cite{Khalid2014,Palomba:ICSME15,Linares:ICSME15}, crash reports from the field in the form of stack traces, or reports in open source issue tracking systems to detect bugs in their apps. In each of these cases, the bug/crash reports are typically lacking in information~\cite{Chen:icse2014,23Joorabchi:ESE13}, containing only a stack trace, overly detailed logs or loosely structured natural language (NL) information regarding the crash \cite{11Bettenburg:MSR08}.  This is not surprising as previous studies showed that information, which is most useful for a developer resolving a bug report (e.g., reproduction steps, stack traces and test cases), is often the most difficult information for reporters to provide \cite{4Joorabchi:MSR14}.  Furthermore, the absence of this information is a major cause of developers failing to reproduce bug/crash reports \cite{3Bettenburg:FSE08}. In addition to the quality of the reports, some other factors specific to Android apps such as hardware and software fragmentation~\cite{android-fragmentation}, API instability and fault-proneness\cite{Linares:FSE13,Bavota:TSE15}, the event-driven nature of Android apps, gesture-based interaction, sensor interfaces, and the possibility of multiple contextual states (e.g., wifi/GPS on/off) make the process of detecting, reporting, and reproducing crashes challenging.

Motivated by these current issues developers face regarding mobile application crashes, we designed and implemented \CrashScope, a practical system that automatically discovers, reports, and reproduces crashes for Android applications.  \CrashScopesp explores a given app using a systematic input generation algorithm and produces expressive crash reports with explicit steps for reproduction in an easily readable natural language format.  This approach requires only an \texttt{.apk} file and an Android emulator or device to operate and requires no instrumentation of the subject apps or the Android OS.  The entirety of the \CrashScopesp workflow is completely automated, requiring no developer intervention, other than reading produced reports. Our systematic execution includes different exploration strategies, aimed at eliciting crashes from Android apps, which include automatic text generation capabilities based on the context of allowable characters for text entry fields, and targeted testing of contextual features, such as the orientation of the device, wireless interfaces, and sensors.  We specifically tailored these features to test the common causes of app crashes as identified by previous studies~\cite{Zaeem:ICST2014,Liang:ICMCN2014,Chandra:GetMobile2015}. During execution, \CrashScopesp captures detailed information about the subject app, such as the inputs sent to the device, screenshots and GUI information, exceptions, and crash information.  This information is then translated into detailed crash reports and replayable scripts, for any encountered crash.

\noindent
This paper makes the following noteworthy contributions:

1) We design and implement a practical and automatic approach for discovering, reporting, and reproducing Android application crashes, called \CrashScope.  To the best of the author's knowledge, this is the first approach that is able to generate expressive, detailed crash reports for mobile apps, including screenshots and augmented NL reproduction steps, in a completely automatic fashion. \CrashScopesp is also one of the only available fully-automated Android testing approaches that is practical from a developers' perspective, requiring no instrumentation of the subject apps or OS.  Our approach builds upon prior research in automated input generation for mobile apps, and implements several exploration strategies, informed by lightweight static analysis that are able to effectively detect crashes and exceptions;

2) We perform a detailed evaluation of the crash detection abilities of \CrashScopesp on 61 Android apps as compared to five state-of-the-art Android input generation tools: Dynodroid\cite{Machiry:FSE13},  Gui-Ripper\cite{Amalfitano:IEEE14}, PUMA\cite{Hao:Mobisys14}, A$^3$E\cite{Azim:OOPSLA13}, and Monkey~\cite{android-monkey}. 
Our results show that \CrashScopesp performs nearly as well as current tools in terms of detecting crashes, while automatically generating detailed reports and replayable scripts;

3) We design and carry out a user study evaluating the \textit{reproducibility} and \textit{readability} of our automatically generated bug reports through comparison to human written crash reports for eight open source apps. The results indicate that \CrashScopesp reports offer more detail, while being at least as useful as the human written reports;

4) We make our experimental dataset and crash reports available in our online appendix\cite{appendix}.

\vspace{-0.15cm}
\section{Related Work \& Motivation}
\label{sec:related}
\vspace{-0.15cm}
In this section, we overview the current landscape of automated testing and input generation tools for Android, discussing limitations of these approaches while illustrating \CrashScopes novelty in context. Several approaches for detecting and reproducing crashes are available in literature \cite{Csallner:SPE2004,Pacheco:ECOOP2005,Csallner:TSEM2008,Pacheco:ICSE2007,Lohman:ICAC2005,Dang:ICSE2012,42Wu:ISSTA2014,8Zhou:ICSE12,Zeller:2005,Zeller:FSE2002,Yoo:TSEM2013,Seo:ASE2012,Liu:FSE2013,46Kim:DSN2011,26Jin:ISSTA13,18Jin:ICSE12};  however, we forgo discussion of these approaches, as they are not presented in the context of mobile apps, and hence do not consider the unique associated challenges.

\vspace{-0.15cm}
\subsection{Input Generation \& Testing Tools for Mobile Apps}
\vspace{-0.1cm}

\begin{table*}[tb]
\vspace{-0.6cm}
\centering
\caption{An Overview of features in automated testing approaches for mobile applications}
\vspace{-0.3cm}
\label{fig:android_approaches}
\scriptsize
\begin{tabular}{| p{65pt} | p{50pt} | p{70pt} | p{69pt} | p{37pt} | p{42pt} | p{39pt} | p{44pt} | p{45pt} |}
\hline
\textbf{Tool Name}& \textbf{Instrumentation} & \textbf{GUI Exploration} & \textbf{Types of Events} & \textbf{Crash Resilient}  & \textbf{Replayable Test Cases} & \textbf{NL Crash Reports} & \textbf{Emulators, Devices} \\ \hline

Dynodroid\cite{Machiry:FSE13}  & Yes & Guided/Random & System, GUI, Text  & Yes & No & No & No  \\ \hline

EvoDroid \cite{Mahmood:FSE14}  & No & System/Evo & GUI & No & No & No  & N/A \\ \hline

\textit{Android}Ripper \cite{Amalfitano:ASE12} & Yes & Systematic & GUI, Text & No & No & No  & N/A \\ \hline 

MobiGUItar \cite{Amalfitano:IEEE14}  & Yes & Model-Based & GUI, Text & No & Yes & No  & N/A\\ \hline

A$^3$E Depth-First \cite{Azim:OOPSLA13}  & Yes & Systematic  & GUI  & No & No & No  & Yes \\ \hline

A$^3$E Targeted\cite{Azim:OOPSLA13}  & Yes & Model-Based  & GUI & No & No & No &  Yes \\ \hline

Swifthand \cite{Choi:OOPSLA13}  & Yes & Model-Based & GUI, Text & N/A & No & No  & Yes   \\ \hline

PUMA \cite{Hao:Mobisys14}  & Yes & Programmable & System, GUI, Text & N/A & No & No  & Yes \\ \hline

ACTEve \cite{Anand:FSE12}  & Yes & Systematic & GUI & N/A & No & No  & Yes  \\ \hline

VANARSena \cite{Ravindranath:Mobisys2014}  & Yes & Random & System, GUI, Text & Yes & Yes & No  & N/A  \\ \hline

Thor \cite{Adamsen:ISSTA2015}  & Yes & Test Cases & Test Case Events & N/A & N/A & No  & No  \\ \hline

QUANTUM \cite{Zaeem:ICST2014}  & Yes & Model-Based & System, GUI & N/A & Yes & No & N/A  \\ \hline

AppDoctor \cite{Hu:ESYS14} & Yes & Multiple & System, GUI$^2$, Text & Yes & Yes & No  & N/A  \\ \hline

ORBIT \cite{Yang:FASE13}  & No & Model-Based & GUI & N/A & No & No  & N/A  \\ \hline

SPAG-C \cite{Lin:TSE14}  & No & Record/Replay & GUI & N/A & N/A & No  & No  \\ \hline

JPF-Android \cite{vanderMerwe:SEN2014}  & No & Scripting & GUI & N/A & Yes & No  & N/A  \\ \hline

MonkeyLab\cite{Linares:MSR15} & No & Model-based & GUI, Text & No & Yes & No  & Yes \\ \hline

CrashDroid \cite{White:ICPC15} & No & Manual Rec/Replay & GUI, Text & Manual & Yes & Yes  &  Yes \\ \hline

SIG-Droid \cite{Mirzaei:ISSRE15} & No & Symbolic & GUI, Text & N/A & Yes & No & N/A  \\ \hline

\textbf{CrashScope}  & \textbf{No} & \textbf{Systematic} & \textbf{GUI, Text, System} & \textbf{Yes} & \textbf{Yes} & \textbf{Yes} & \textbf{Yes}  \\ \hline
\end{tabular} 
\vspace{-0.7cm}
\end{table*}

Approaches for automated input generation can be broadly grouped into three major categories \cite{Choudhary:ASE15}: \textit{random-based input generation}\cite{Machiry:FSE13,android-monkey,intent-fuzzer,Sasnauskas:WODA14,Ye:MoMM13}, \textit{systematic input generation}\cite{Azim:OOPSLA13,Anand:FSE12,Amalfitano:ASE12}, and \textit{model-based testing}\cite{Amalfitano:ASE12,Yang:FASE13,Azim:OOPSLA13,Choi:OOPSLA13,Hao:Mobisys14,Zaeem:ICST2014}. We outline some features of several mobile testing and input generation tools in Table \ref{fig:android_approaches}.

\textbf{Random-Based Input Generation} techniques rely on choosing arbitrary GUI or contextual events; they can adapt such a strategy through biasing the selection with the frequency of past events and purposefully select those previously un-exercised. Dynodroid \cite{Machiry:FSE13} introduces such an approach by maintaining a history of event execution frequencies in a context-sensitive manner that tracks relevant events throughout the execution.   Intent Fuzzer \cite{Sasnauskas:WODA14} generates intents to test apps by taking into consideration information extracted in an offline static analysis phase. However, it does not easily scale for large apps due to the number of paths that need to be stored in memory for processing, and only tests app intents. VanarSena \cite{Ravindranath:Mobisys2014} is a tool for Windows Phone that instruments app binaries in order to test apps for externally-inducible faults using random exploration and the injection of adverse contextual conditions.  VanarSena  is capable of generating crash reports containing a stack trace and exception. Compared to VanarSena, \CrashScopesp does not require any type of instrumentation and is able to generate more detailed crash reports that include screenshots, natural language reproduction steps, and replayable scripts.

\textbf{Systematic Input Generation} approaches execute input events according to a pre-defined heuristic such as Breadth or Depth First Search (BFS/DFS). For instance, \textit{Android}Ripper dynamically analyzes an app's GUI with the aim of obtaining a fireable sequence of events.  It then extracts task lists according to the GUI hierarchy of an app and systematically executes the events in the generated lists. A$^3$E \cite{Azim:OOPSLA13} leverages static bytecode, taint-style, dataflow analysis in order to construct a high-level control flow-graph that captures allowable transitions between app screens (activities). The tool implements two exploration strategies based on this graph, a simple DFS, and targeted exploration.  ACTEve \cite{Anand:FSE12} is a  concolic-based testing approach for smartphone apps that symbolically tracks events from the point where they originate to the point where they are handled. The tool is similar in efficacy to concolic testing while avoiding the path-explosion problem, but requires Soot for instrumentation \cite{soot}. 

\textbf{Model-Based Input Generation} approaches attempt to build detailed models (e.g., capturing relevant screen and event flows) of apps in order to be able to explore them more effectively.  MobiGUItar \cite{Amalfitano:IEEE14} is the evolution of AndroidRipper and models the states of an Android app's GUI, executing feasible events by observing and updating the current GUI state of an app. Swifthand \cite{Choi:OOPSLA13} uses active learning to infer a model of an app and uses the learned model to generate inputs that drive the execution of the app towards unexplored states. Additionally, this approach aims to minimize app restarts, due to restart overhead, by exploring all screens that can be reached from the initial state of the app first. QUANTUM \cite{Zaeem:ICST2014} represents an approach for automatically generating GUI-based app-agnostic oracles for the detection of defects in Android applications. The tool implements a set of app-agnostic features, namely: rotation, killing and restarting, pausing and resuming, and back button. It is able to present a visual representation of the app to the developer for verification after exercising the previous features at targeted locations during testing. While QUANTUM can not produce detailed crash reports, it can produce replayable Junit/Robotium test cases. ORBIT \cite{Yang:FASE13} uses static analysis to extract GUI-components in the layout files and locate any handlers or event listeners in the source files in order to extract possible fireable actions. Then, it uses dynamic GUI crawling to construct a model of an app. MonkeyLab \cite{Linares:MSR15} is an approach that mines application usages to extract \textit{n}-grams representing the GUI and usage models.  MonkeyLab is capable of generating unseen and actionable sequences of events that achieve orthogonal code coverage compared to the application usages mined to generate the model.  Tonella et. al. \cite{Tonella:ICSE2014} also applied interpolated \textit{n}-grams to model based testing, but not in the context of mobile apps.

\textbf{Other Approaches in Mobile Testing and Bug Reporting} encompass emerging work that attempts to solve specific problems related to testing and bug reporting in a mobile context.  PUMA \cite{Hao:Mobisys14} exposes high-level events for which users can define handlers, in order to create a programmable UI-Automation framework for which developers can implement their own exploration strategies. This system relies on the instrumentation of app binaries and implements a basic random input generation approach.  Thor \cite{Adamsen:ISSTA2015} relies on existing test cases for an app and systematically triggers adverse contextual conditions during the execution of the sequences in given test case. AppDoctor \cite{Hu:ESYS14} takes an ``approximate execution" approach by using a side loaded instrumentation app to hook directly into event handlers that are associated with various application GUI-components.  While this approach does offer the ability to replay the bugs and presents the developer with stack traces, it must replay crash traces to prune false-positives and it does not offer highly detailed and expressive reports as \CrashScopesp does. SIG-Droid~\cite{Mirzaei:ISSRE15} infers test inputs by relying on symbolic execution, and combining inputs with a GUI model extracted automatically from the source code. SPAG-C \cite{Lin:TSE14} is aimed at determining whether or not a given device is capable of correctly displaying an app that is assumed to be working correctly. This approach combines Sikuli, for UI Automation, and Android screencast for capturing the frame buffer of a device.  JPF-Android \cite{vanderMerwe:SEN2014} verifies Android apps by running android code on the JVM using a collection of different event sequences and then detecting when certain errors occur using Java PathFinder (JPF).  This approach introduces overhead due to the stack manipulation, listeners and logging used by JPF, and is severely limited in terms of the types of events it can properly execute. Evodroid \cite{Mahmood:FSE14} expands upon a typical systematic strategy and leverages evolutionary algorithms that use Interface and Call-Graph models. 

\indent The motivation for \CrashScopesp stems in part from our previous work in mobile bug and crash reporting \cite{White:ICPC15,Moran:FSE15,Moran:ICSE16}.  \CrashDroid \cite{White:ICPC15} is capable of translating a call stack from an app crash into expressive steps to reproduce a bug. However, it has two noteworthy limitations: first, it is not able to uncover crashes automatically, but instead relies on pre-existing stack traces; second, the tool requires expensive collection of manual traces from real users that need to be annotated using natural language descriptions, which are needed to construct the bug reports and replayable scenarios.  FUSION \cite{Moran:FSE15,Moran:FSESRC2015,Moran:ICSE16} is a bug reporting mechanism that leverages program analysis to facilitate users creating expressive bug reports.  However, this approach requires users to create bug reports for functional problems in an app.  \CrashScopesp automates the entire crash discovery process and adds information regarding the contextual state of the app at each reproduction step, resulting in a tool that automates a typically expensive maintenance process for developers.
	
\vspace{-0.15cm}
\subsection{Previous Studies on Mobile App Bug/Crashes}
\vspace{-0.1cm}
	Motivating factors from mobile app bug/crash studies aided us in designing \CrashScope. Two studies stand out in terms of providing information to drive design decisions for our approach. First, Ravindranath \emph{et al.} \cite{Ravindranath:Mobisys2014} conducted a study of 25 million real-world crash reports collected from Windows Phone users ``in the wild" by the ``Windows Phone Error Reporting System" (WPER).  Although this study was conducted regarding crashes from a different mobile OS, several of the findings reported in this study are relevant in the context of Android, due to platform similarities: 1) a small number of root causes cover a majority of the crashes examined;  2) many crashes can be mapped to well-defined externally inducible faults, for example, HTTP errors caused by network connectivity issues;  3) the dominant root causes can affect many different user execution paths in an app.  The most salient piece of information that can be gleaned from the study and applied in the design of \CrashScopesp is the following: \textit{An effective crash discovery tool must be able to test different contextual states in a targeted manner, while remaining resilient to encountered crashes so as to uncover crashes present in different program event-sequence paths.} We explain how \CrashScopesp achieves targeted testing of contextual states using program analysis in Sec. \ref{sec:approach}.
	
	In addition, Zaeem \emph{et al.} \cite{Zaeem:ICST2014} conducted a bug study on 106 bugs drawn from 13 open-source Android applications, with the goal of identifying opportunities for automatically generating test cases that include oracles. Most notably, the results of this study were formulated as a categorization of different Android app bugs.  Specifically, these categorizations were grouped into three headings: Basic Oracles, App-Agnostic Oracles, and App-Specific Oracles.  \CrashScopesp uses the well-defined oracles of uncaught exceptions and app crashes to detect faults; however, some of the bug categorizations in this study are useful in triggering these, specifically the app-agnostic categorizations of \textit{Rotation}, \textit{Activity Life-Cycle}, and \textit{Gestures}.  Specifically, we implemented a targeted (i.e. localized) version of the double-rotation feature \cite{Zaeem:ICST2014}.
	
\begin{figure*}[tb]
\vspace{-0.9cm}
\centering
\includegraphics[width=\linewidth]{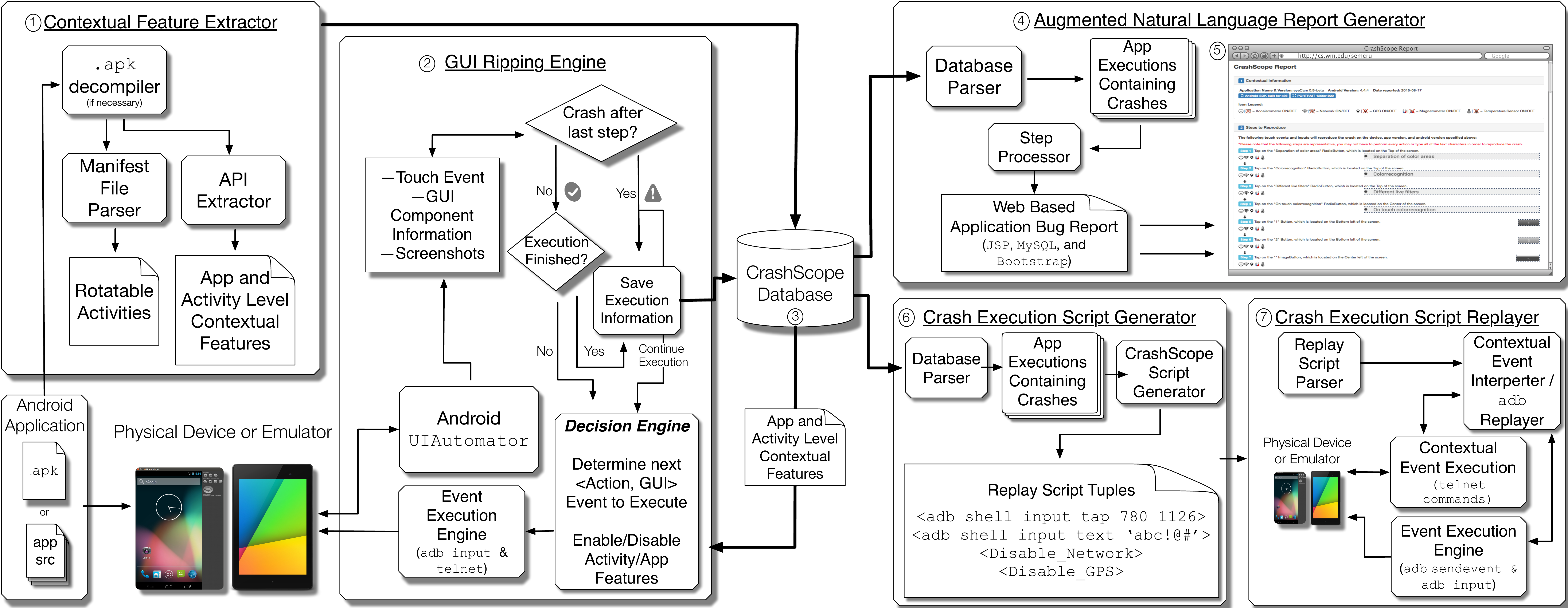}
\vspace{-0.7cm}
\caption{Overview of CrashScope Workflow}
\label{Design}
\vspace{-0.7cm}
\end{figure*}

\vspace{-0.15cm}
\subsection{Limitations of Mobile Testing Approaches}
\vspace{-0.1cm}
\label{subsec:limitations}

While significant progress has been made in the area of testing and automatically generating inputs for mobile applications, the available tools generally exhibit some noteworthy limitations that inspired the development of \CrashScope:
	
\begin{itemize}
\item{Previous approaches lack the ability to provide detailed, easy-to-understand testing results for faults discovered during automatic input generation, leaving the developer to sort through and comprehend stack traces, log files, and non-expressive event sequences \cite{Choudhary:ASE15}};
\item{Most approaches for automated input generation are not practical for developers to use, typically due to instrumentation or difficult setup procedures.  This is affirmed by the fact developers typically prefer manual over automated testing approaches \cite{Kochhar:ICST15,23Joorabchi:ESE13,Linares:ICSME15}}. As we show, instrumentation can contribute to a higher than necessary developer effort in parsing results from automated approaches, because developers must sift through ``false positive" crashes caused by instrumentation. 
\item{Few approaches combine different strategies and features for testing apps through supporting different methodologies for user text input and testing of contextual states (e.g., wifi on/off) in a single holistic approach.}
\end{itemize}

These shortcomings contribute to the low adoption rate of automated testing approaches by mobile developers. In the next section of this paper, we clearly describe how \CrashScopes design addresses these current limitations in automated mobile input generation and testing tools.

\vspace{-0.15cm}
\section{CrashScope Design}
\label{sec:approach}
\vspace{-0.15cm}
	In this section, we first describe \CrashScopes novelty by illustrating how it addresses the limitations discussed in the previous section.  We then give an overview of the \CrashScopes workflow, and the salient features in detail.

	\CrashScopesp addresses the general limitations of existing tools. First, no other automated testing approach, is able to automatically generate \textit{expressive} bug reports (and replayable scripts) for exceptions and crashes discovered by automated input generation for mobile apps. \CrashScopesp accurately detects crashes and is able to generate \textit{easily readable and detailed} reports without any developer intervention.  Second, \CrashScopesp is a practical tool, requiring only an \texttt{.apk} file and an instantiated emulator \textit{or} physical device running Android 4.3 and newer, which constitutes 55\% of the current Android OS install base\cite{android-platform}. Operating on emulators \CrashScopesp is able to parallelize testing on multiple emulators with different specifications, versions of Android, and screen sizes, mitigating a major challenge in app development\cite{23Joorabchi:ESE13}.  Third, inspired by existing approaches \cite{Ravindranath:Mobisys2014, Zaeem:ICST2014, Adamsen:ISSTA2015} \CrashScopesp is able to explore an app through automated input generation while testing varying contextual states.  We extend previous  context aware testing techniques by leveraging static analysis to extract targeted locations for testing apps in different contextual states.    Finally, our approach is \textit{app-crash-resilient}; it can detect a crash and continue testing the unvisited components and states of the GUI after handling the crash.
 
	The overall workflow of \CrashScopesp is illustrated in Figure \ref{Design}. Let us consider the 31C3 Schedule app \cite{schedule} as a running example to explain the \CrashScopesp  workflow; then, we will discuss the salient features in detail.  The first step in running \CrashScopesp is to obtain the source code of the app, either directly or through decompilation, and detect Activities (by means of static analysis) that are related to contextual features (Figure \ref{Design}-\circled{1}) in order to target the testing of such features.  In other words, \CrashScopesp will only test certain contextual app features  (e.g., wifi off) if it finds instances where they are implemented in the source code.  In the case of 31C3 Schedule, the first activity (screen) of the app makes use of network connectivity, so this screen would be marked as implementing this feature. More details about the contextual features detection are provided in Sec. \ref{subsec:context-features}.

	Next, the \textit{GUI Ripping Engine} (Figure \ref{Design}-\circled{2}) systematically executes the app using various strategies (Section \ref{subsec:strategies}), including enabling and disabling the contextual features (if run on an emulator) at the Activities of the app identified previously.  If during the execution, uncaught exceptions are thrown, or the app crashes,  dynamic execution information is saved to the \CrashScope's database (Figure \ref{Design}-\circled{3}), including detailed information regarding each event performed during the systematic exploration.  In the case of 31C3 Schedule, if systematic execution is continued from the first screen when the network is disabled, a crash occurs.  This is because the differing contextual condition exposes a state of the app that would not be otherwise seen.
	
	After the execution data has been saved to the \CrashScopesp database, the \textit{Natural Language Report Generator} (Figure \ref{Design}-\circled{4}, Section \ref{subsec:reports}) parses the database and processes the information for each step of all executions that ended in a crash, generating an HTML based natural language crash report with expressive steps for reproduction (Figure \ref{Design}-\circled{5}).  In addition, the \textit{Crash Script Generator} (Figure \ref{Design}-\circled{6}, Section \ref{subsec:scripts}) parses the database and extracts the relevant information for each step in a crashing execution in order to create a replayable script containing \texttt{adb input}  commands and markers for contextual state changes.  The \textit{Script Replayer} (Figure \ref{Design}-\circled{7}, Section \ref{subsec:scripts}) is able to replay these scripts by executing the sequence of \texttt{adb input} commands and interpreting the contextual state change signals, in order to reproduce the crash.  In the case of the 31C3 Schedule app, this involves turning off the network connection and trying to interact with one of the app menu headers.

\vspace{-0.15cm}
\subsection{Extracting Activity and App-Level Contextual Features}
\label{subsec:context-features}
\vspace{-0.1cm}

	\CrashScopesp uses Abstract Syntax Tree (AST) based analysis to extract the API-call chains that are involved in invocations of contextual features. In particular, it detects Android API calls related to network connectivity and sensors (i.e., Accelerometer, Magnetometer, Temperature Sensor, and GPS). Because the API calls might not be executed directly by an Activity, \CrashScopesp performs a call-graph analysis to extract paths ending in a method invoking a contextual API.  Because certain API calls may not be traceable through a back-propagated call-chain (e.g., sensor or network implemented as a service), \CrashScopesp employs two granularities for testing contextual features: activity (screen-) level and app-level.  If a particular API call related to one of the contextual features above is able to be traced back to an Activity, then that feature is later tested at the Activity level  (i.e., the contextual feature is enabled or disabled when the corresponding Activity is in foreground).  If the feature is not able to be linked to an Activity, then the feature is tested at the level of the entire app (i.e., the contextual feature is enabled or disabled at the beginning of the app's execution).  To obtain the Activities that are rotatable, \CrashScopesp parses the AndroidManifest.xml, where rotatable activities must be declared.  During testing, if a rotatable activity is encountered while exploring an app, then the screen is rotated from portrait to landscape and back again before any GUI interactions occur to test for proper implementation of the corresponding rotation event-handlers.

\vspace{-0.15cm}
\subsection{Exploration of Apps \& Crash Detection}
\vspace{-0.1cm}

	To explore an app, \CrashScopesp dynamically extracts the GUI hierarchy of each app screen visited during the exploration and identifies the clickable and long-clickable components to execute, as well as available components for text inputs (e.g., EditText boxes). The (long-) clickable components are added to a working list to assure that all the clickable components are executed systematically.   \CrashScopesp executes each possible event (i.e., action on an available GUI component) on the current screen according to the GUI hierarchy. If text entry fields are available in a particular app screen, then each text box is filled in before each (long-) clickable component on the screen is exercised.  Currently, our \textit{Ripping Engine} supports the \textit{tap}, \textit{long-tap}, and \textit{type} events.
	
	Text entry from the user is a major part of functionality in many Android apps, therefore, \CrashScope's \textit{GUI Ripping Engine} employs a unique text input generation mechanism.  \CrashScopesp detects the type of text expected (e.g., numbers) by a text field, by querying the keyboard type associated to the text field \cite{android-keyboard}. This is done with the  \texttt{adb shell dumpsys input\_method} command. Once the type of expected input is detected, \CrashScopesp employs two strategies to generate text inputs: \textit{expected} and \textit{unexpected}. The \textit{expected} strategy generates a string within the keyboard parameters without any punctuation or special characters, whereas the \textit{unexpected} strategy generates random strings with all of the allowable special characters for a given keyboard type.  The intuition behind this input generation mechanism is to test instances where a developer may have unknowingly set a keyboard that allows certain characters, but does not properly check for these characters in the code, resulting in a fault.  Before the keyboard metadata is read, a \textit{touch} event is executed on the text box to ensure the corresponding keyboard is displayed.
	
   In addition to the text input generation strategies, \CrashScopesp traverses the GUI hierarchy either from the bottom of the hierarchy up or from the top of the hierarchy down.  The rationale for having two such strategies is to generally mimic what a user would do, i.e., executing GUI events without a predefined order. If a transition to another screen is recorded during the exploration, then the GUI-hierarchy of the new screen is detected and the components on the new screen are executed next.  The \textit{GUI Ripping Engine} constructs a graph containing all of the possible transition states and uses the back button to return to previous states after the executable components in a particular branch have been exhausted.  It also keeps a stack of all the yet-to-be visited components.  To detect and capture exceptions, \CrashScopesp filters the logcat for uncaught exceptions related only to the app being tested.  To detect crashes, \CrashScopesp checks for the appearance of the standard Android crash dialog.  If a crash is encountered, the execution information is logged to the database, but because of the transition diagram and stack of unvisited components, execution can continue towards additional remaining program paths without starting the execution from scratch.

\vspace{-0.15cm}
\subsection{Testing Apps in Different Contextual States}
\vspace{-0.1cm}

	When the GUI-Ripping begins, \CrashScopesp first checks for app-level contextual features that should be tested according to the exploration strategy.  Then, the \textit{GUI Ripping Engine} checks if the current Activity is suitable for exercising a particular contextual feature in adverse conditions.  If this is the case, it sets the value of the sensor according to the current strategy.  The testing of contextual features works \textit{only on} emulators using telnet commands associated with standard Android Virtual Devices (AVDs) \cite{avd}.  While the telnet commands do support turning on/off the network for an emulator, they do not support the enabling/disabling of sensors (Accelerometer, Magnetometer, GPS, Temperature Sensor), but it is possible to set the values of these sensors.  Therefore, to test for sensor related features in adverse conditions, the network connection is disabled, and unexpected values are set for the other sensors (GPS, Accelerometer, etc) that would not typically be possible under normal conditions. For instance, to test the GPS in an adverse contextual state, \CrashScopesp sets the value to coordinates that do not represent physical GPS coordinates.

\vspace{-0.15cm}
\subsection{Multiple Execution Strategies}
\label{subsec:strategies}
\vspace{-0.15cm}

	One of \CrashScopes most powerful features is its ability to explore an app according to several different strategies through combinations of its various supported testing features.  These strategies stem from three major feature heuristics: 1) the direction in which to traverse the GUI Hierarchy (top-down or bottom-up), 2) the method by which inputs are generated for user text entry fields (\textit{no text}, \textit{expected text}, \textit{unexpected text}), and finally, 3) enabling or disabling the testing of adverse contextual states (e.g., if an activity is found to have utilize wifi, should it be turned on or off?).  Different combinations of these strategies have the potential to uncover different types of app crashes.  For example, consider the following configuration \texttt{<no\_text, top\_down, enable\_all\_context\_states>}.  According to this strategy, \CrashScopesp will not enter any user text, will exercise the GUI-components in order from the top of the screen to the bottom, and will trigger adverse contextual features in activities where they are detected.  This type of strategy has a high likelihood of uncovering crashes like the one described earlier in C13C Schedule in which the change of contextual state triggers a crash.  However, the \texttt{<unexpected\_text, top\_down, disable\_context\_states>} has a better chance of uncovering crashes related to user input being handled improperly by the app.  By running an app through all 12 combinations of these three feature heuristics in different strategies, \CrashScopesp can effectively test for different types of commonly inducible crashes.  These strategies can also be parallelized by running several strategies for an app concurrently on a group or cloud of emulator instances, further reducing the testing overhead for the developer.

\subsection{Generating Expressive, Natural Language Crash Reports}
\label{subsec:reports}
\begin{figure}
\centering
\includegraphics[width=\columnwidth]{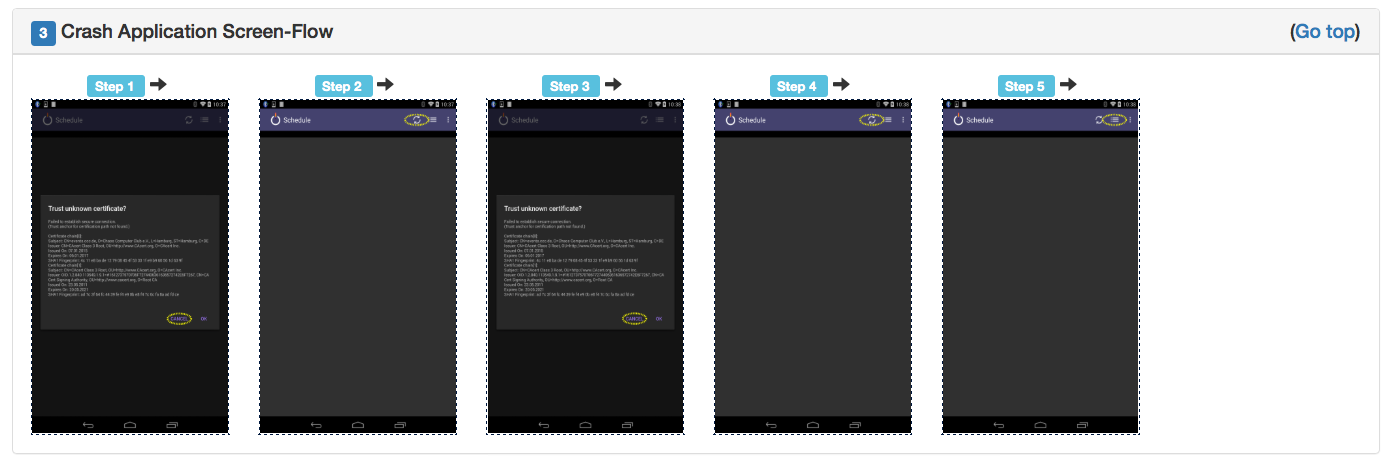}
\vspace{-0.7cm}
\caption{Crash Screen-Flow}
\label{screenflow}
\vspace{-0.1cm}
\end{figure}

\begin{figure}
\vspace{-0.5cm}
\centering
\includegraphics[width=\columnwidth]{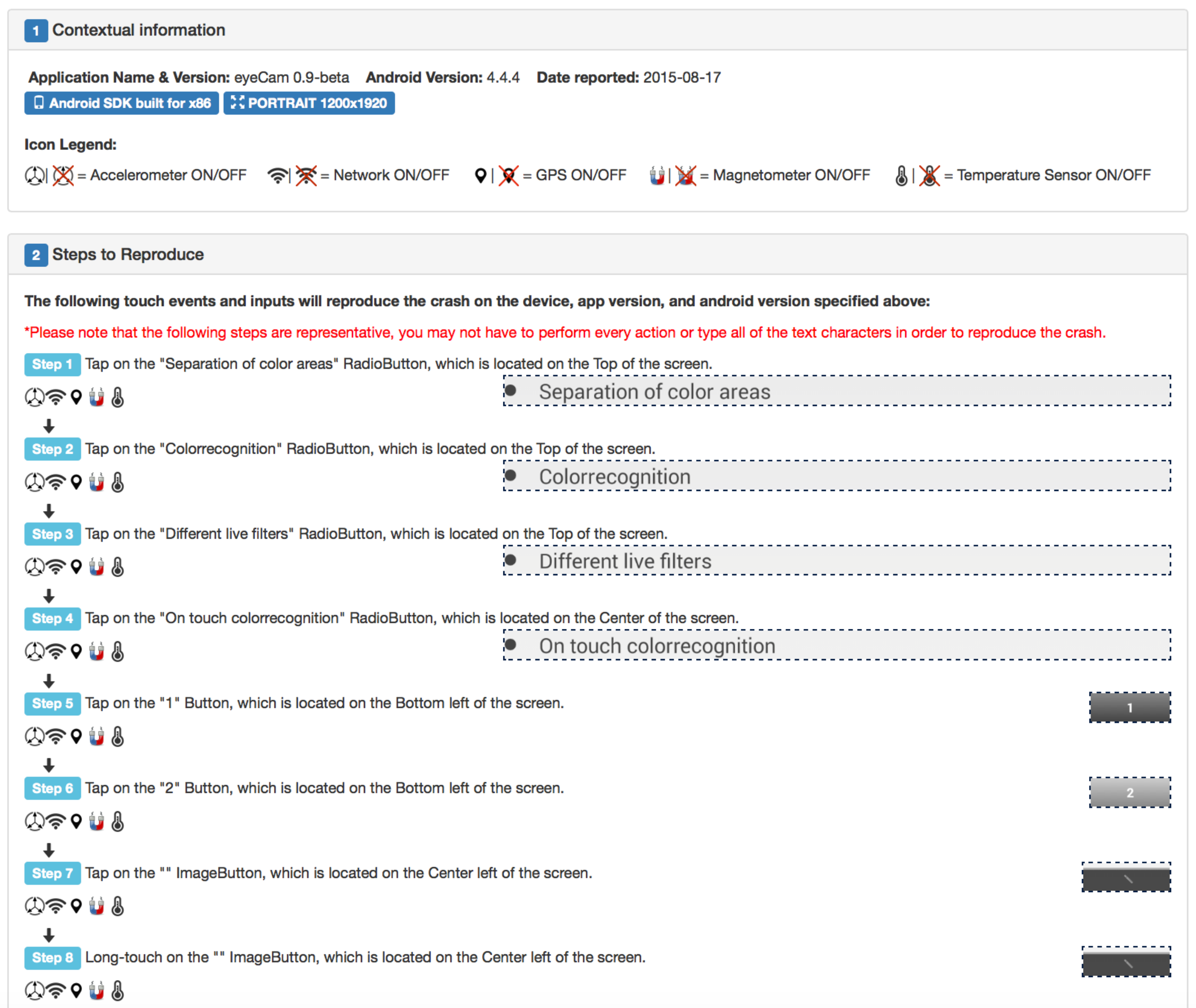}
\vspace{-0.7cm}
\caption{Example of Contextual Information and Reproduction Steps sections in a generated crash report}
\vspace{-0.1cm}
\label{steps}
\end{figure}

	\CrashScopesp generates a Crash Report (Figure \ref{Design}-\circled{5}) that contains four major types of information: 1) general information including the app name and version, the version of the Android OS, a legend of icons that indicate the current contextual state of the app in the reproduction steps, the device, and the screen orientation and resolution when the crash occurred; 2) natural language sentences that describe the steps to reproduce a crash using detailed information about the GUI events and contextual states for each step (Figure \ref{steps}); 3) an app's screen flow that highlights the component interacted with on each screen in the execution scenario for a particular crash (Figure \ref{screenflow}); (4) a pruned stack trace containing only the app exceptions that occurred during execution.
	 
	The natural language reproduction steps are constructed by the \textit{Report Generator} (Figure \ref{Design}-\circled{4}) using the template: 
\begin{quote}
\small	
\texttt{<action>} \textit{on} \texttt{<component text> <component type>}, \textit{which is located on the} \texttt{<relative location>} \textit{of the screen}
\end{quote}

	For the steps that have text entry associated with them, the \texttt{<action>} placeholder is modified into the following: ``\texttt{Type <text input>} on the..." so as to capture any specific text inputs that may trigger a crash. 

\vspace{-0.15cm}
\subsection{Generating \& Replaying Reproduction Scripts}
\label{subsec:scripts}
\vspace{-0.1cm}

	The  \textit{Crash Script Generator} (Figure \ref{Design}-\circled{6}), parses the saved execution information from the \CrashScopesp database and generates replayable scripts containing \texttt{adb input} commands for touch and text inputs and markers for changes in contextual states.  The scripts are generated by parsing the database for all of the GUI events associated with each step in a particular execution.  Then, the coordinates of each component that were recorded during the systematic exploration of the app are parsed and the center coordinates are extrapolated based on each component's size.  These coordinates are used to generate \texttt{adb input} commands to reproduce the GUI event.  This approach relies on our previous work in replaying events of test sequences in Android apps \cite{Linares:ICSE15,Linares:MSR15}. An example of a \CrashScopesp replayable script can be seen in Fig. \ref{Design}-\circled{6}. The scripts can be replayed by the  \textit{Script Replayer} (Fig. \ref{Design}-\circled{7}), which executes the \texttt{adb input} commands, and interprets the state change markers in the script (e.g.,$\langle$Wifi\_OFF$\rangle$) to execute proper \textit{telnet} commands to set states on an emulator.  
	
\vspace{-0.15cm}
\section{Empirical Study 1: Crash Detection Capability}
\label{sec:study1}
\vspace{-0.15cm}
The \textit{goal} of our first study is to evaluate the effectiveness of \CrashScopesp at discovering crashes in Android apps as compared to state-of-the-art approaches for testing mobile apps. The \textit{quality focus} of this first study concerns the fault detection capabilities of \CrashScopesp in terms of locating crashes.  The \textit{context} of this study consists of 61 open-source Android apps previously used to evaluate automated testing approaches in \cite{Choudhary:ASE15}, as well as five approaches for automated input generation (listed in Table \ref{tool-table}). 
We investigated the following research questions (RQs):
\begin{itemize}
\item{\textbf{RQ$_1$}: \textit{What is \CrashScopes effectiveness in terms of detecting application crashes compared to other state-of-the-art Android testing approaches?}}
\item{\textbf{RQ$_2$}: \textit{Does \CrashScopesp detect different crashes compared to the other tools?}}
\item{\textbf{RQ$_3$}: \textit{Are some \CrashScopesp execution strategies more effective at detecting crashes or exceptions than others?}}
\item{\textbf{RQ$_4$}: \textit{Does average application statement coverage correspond to a tool's ability to detect crashes?}}
\end{itemize}

\vspace{-0.15cm}
\subsection{Methodology}
\vspace{-0.1cm}

	In order to compare \CrashScopesp against other state-of-the-art automated input generation tools for Android, we utilized a subset of subject apps and tools available in the Androtest testing suite \cite{Choudhary:ASE15, androtest}.  We chose to perform this study on a subset of the tools offered by the Androtest artifact due to runtime issues, namely, some tools would not run consistently on the set of provided subject apps (e.g., the tools would launch an emulator but not the app), causing inconsistent results we chose to exclude.  However, when contacted, the authors of the tool were helpful in supporting us.  We believe the tools tested against constitute a diverse representation of the publicly available Android testing tools.  The Androtest suite contains 68 subject applications for testing; however, when recompiling the applications to run the tools and extract the apps from the VM to run with \CrashScope, seven of the subject apps failed to compile with the instrumentation necessary to gather code-coverage results.  Therefore, each tool in the suite was allowed to run for one hour for each of the remaining 61 subject apps, five times, whereas we ran all 12 combinations of the \CrashScopesp strategies once on each of these apps. It is worth noting that the execution of tools in the Androtest suite (except for Android monkey) can not be controlled by a criteria such as maximum number of events. 
		
\begin{table}[t]
\vspace{-0.6cm}
 \centering
 \caption{Tools used in the comparative fault finding study}
 \label{tool-table}
\vspace{-0cm}
 \begin{tabular}{  |c | c | c| }
  \hline                       
  Tool Name & Android Version & Tool Type \\ \hline
  Monkey & any & Random \\
  A$^3$E Depth-First & any & Systematic\\\ 
  GUIRipper & any & Model-Based\\
  Dynodroid & v2.3 & Random-Based\\
  PUMA & v4.1+ & Random-Based\\
  \hline  
\end{tabular}
\vspace{0.35cm}
\end{table}
		
	 In the Androtest VMs, each tool ran on its required Android version, for \CrashScopesp each subject application was run on an emulator with a 1200x1920 display resolution, 2GB of RAM, a 200 MB Virtual sdcard, and Android version 4.4.2 KitKat.  We ran the tools listed in Table \ref{tool-table}, except Monkey, using Vagrant\cite{vagrant} and VirtualBox\cite{virtualbox}.  The Monkey tool was run for 100-700 event sequences (in 100 event deltas for seven total configurations) on an emulator with the same settings as above with a two-second delay between events, discarding trackball events. Trackball events were discarded to facilitate a fair comparison to the supported events and hardware of the other testing tools. Each of these seven configurations was executed five times for each of the 61 subject apps, and every execution was instantiated with a different random seed \cite{android-monkey}.  While Monkey is an available tool in Androtest, the authors of the tool chose to set no delay between events, meaning the number of events monkey executed over the course of 1 hour far exceeds the number of events generated by the other tools, which would have resulted in a biased comparison to \CrashScopesp and the other automated testing tools. To facilitate a fair comparison, we chose to limit the number of events thrown by Android monkey to a range (100-700 events) that corresponds to the average number of events invoked by other tools.  In order to give a complete picture of the effectiveness of \CrashScopesp as compared to the other tools, we report data on both the statement coverage of the tools as well as crashes detected by each tool.  Each of the subject applications in the Androtest suite was instrumented with the Emma code coverage tool \cite{emma}, and we used this instrumentation to collect statement coverage data for each of the apps. Due to space limitations, we report the cumulative coverage for all of the strategies and runs of each tool with a full dataset of detailed statistics available in our replication package in the online appendix \cite{appendix}.\\
	\indent The underlying purpose of this study is to compare the crash detection capabilities of each of these tools and answer \textbf{RQ$_1$}. However, we cannot make this comparison in a straightforward manner.  \CrashScopesp is able to accurately detect app crashes by detecting the standard Android dialog for exposing a crash (e.g., a text box containing the phrase ``application\_name \textit{has stopped}").  However, because the other analyzed tools do not support identifying crashes at runtime, there is no reliable automated manner to extract instances where the application crashed purely from the \texttt{logcat}\cite{logcat}. To obtain an approximation of the crashes detected by these tools, we parsed the \texttt{logcat} files generated for each tool in the Androtest VMs. Then, we isolated instances where exceptions occurred containing the \texttt{FATAL EXCEPTION} key marker, which were also associated with the process id (pid) of the app running during the \texttt{logcat} collection.  While this filters out unwanted exceptions from the OS and other processes, unfortunately, it does not guarantee that the exceptions signify a crash caused by incorrect application logic.  This could signify, among other things, a crash caused by the instrumentation of the controlling tool.  Therefore, in order to conduct a consistent comparison to \CrashScope, the authors manually inspected the instances of fatal exception stack traces returned by the \texttt{logcat} parsing, discarding duplicates and those caused by instrumentation problems, and we report the crash results of the other tools from this pruned list.  A full result set with both full and pruned \texttt{logcat} traces is available in our online appendix \cite{appendix}.  The issues encountered when parsing the results from these other tools further highlight \CrashScopes utility, and the need for an automatic tool that can accurately detect and in turn effectively report crashes in mobile apps.	

\begin{table}[t]
\vspace{-0.6cm}
\centering
\caption{Unique Crashes Discovered with Instr. Crashes in parenthses}
\label{tab:crashes}
\setlength{\tabcolsep}{0.6em}
\def\arraystretch{0.7}
\vspace{-0.3cm}
\begin{tabular}{| p{37pt} | l | p{25pt} | p{25pt} | l | p{25pt} | p{25pt} |}
\hline
            App     & A$^3$E & GUI- Ripper & Dyno- droid & PUMA & Monkey (All) &Crash- Scope  \\ \hline
A2DP Vol          & 1   & 0         & 0         & 0  & 0 & 0          \\ \hline     
aagtl         & 0   & 0         & 1         & 0  & 1 & 0          \\ \hline       
Amazed & 0   & 0         & 0        & 0  & 1 & 0          \\ \hline
HNDroid          & 1   & 1         & 1        & 2  & 1 & 1          \\ \hline
BatteryDog & 0   & 0         & 1        & 0  & 1 & 0          \\ \hline
Soundboard       & 0   & 1         & 0         & 0  & 0 & 0          \\ \hline
AKA & 0   & 0         & 0        & 0  & 1 & 0          \\ \hline
Bites       & 0   & 0         & 0         & 0  & 1 & 0          \\ \hline
Yahtzee          & 1   & 0         & 0         & 0    &0& 1          \\ \hline
ADSDroid         & 1   & 1         & 1         & 1    &1& 1         \\ \hline
PassMaker & 1   & 0         & 0         & 0    &1& 1          \\ \hline
BlinkBattery & 0   & 0         & 0        & 0  & 1 & 0          \\ \hline
D\&C & 0   & 0         & 0        & 0  & 1 & 0          \\ \hline
Photostream      & 1   & 1         & 1         & 1    &1& 0          \\ \hline
AlarmKlock       & 0   & 0         & 1         & 0    &0& 0          \\ \hline
Sanity           & 1   & 1         & 0         & 0    &0& 0          \\ \hline
MyExpenses       & 0   & 0         & 1         & 0    &0& 0          \\ \hline
Zooborns         & 0   & 0         & 0         & 0    &0& 2          \\ \hline
ACal             & 1   & 2         & 2         & 0    &1& 1          \\ \hline
Hotdeath         & 0   & 2         & 0         & 0    &0& 1          \\ \hline
\textbf{Total}  & \textbf{8 (21)}   & \textbf{9 (5)}  & \textbf{9 (6)}  & \textbf{4 (0)}    & \textbf{12 (1)}  &  \textbf{8 (0)} \\ \hline
\end{tabular}
\vspace{-0.1cm}
\end{table}

\vspace{-0.15cm}
\subsection{Results \& Discussion}
\vspace{-0.1cm}
 
 	Table \ref{tab:crashes} shows the aggregated crash discovery results of each tool over their various runs.  This table reports unique crashes (as signified by differing stack traces not caused by app instrumentation) detected by the various approaches, and only includes those apps for which crashes were discovered. For tools other than \CrashScope, we also report crashes (in parentheses) that were caused by instrumentation frameworks (e.g. troyd, Android intsr., junit, Emma), as these represent ``false positive" crashes uncovered by the tools. The results highlight four key results.  The first observable result is that \CrashScopesp is about as effective in terms of number of crashes detected, while also providing detailed bug reports. CrashScope discovered fewer crashes compared to Monkey due to the large number of events that this tool is capable of producing.  However, it should be noted that Monkey is not able to generate replayable scripts or reports, severely limiting its usefulness form a developers perspective.  \CrashScopesp was able to discover about as many crashes as A$^3$E, GUI-Ripper, and Dynodroid, more than PUMA,  without any false positives caused by instrumentation of the app or system.  Therefore, we answer \textbf{RQ$_1$} as follows: \textbf{CrashScope is about as effective at detecting crashes as the other tools.  Furthermore, our approach reduces burden on developers by reducing the number of ``false" crashes caused by instrumentation and providing detailed crash reports}.
		
 \begin{figure}[t]
 \vspace{-0.6cm}
\centering
\includegraphics[width=0.6\columnwidth]{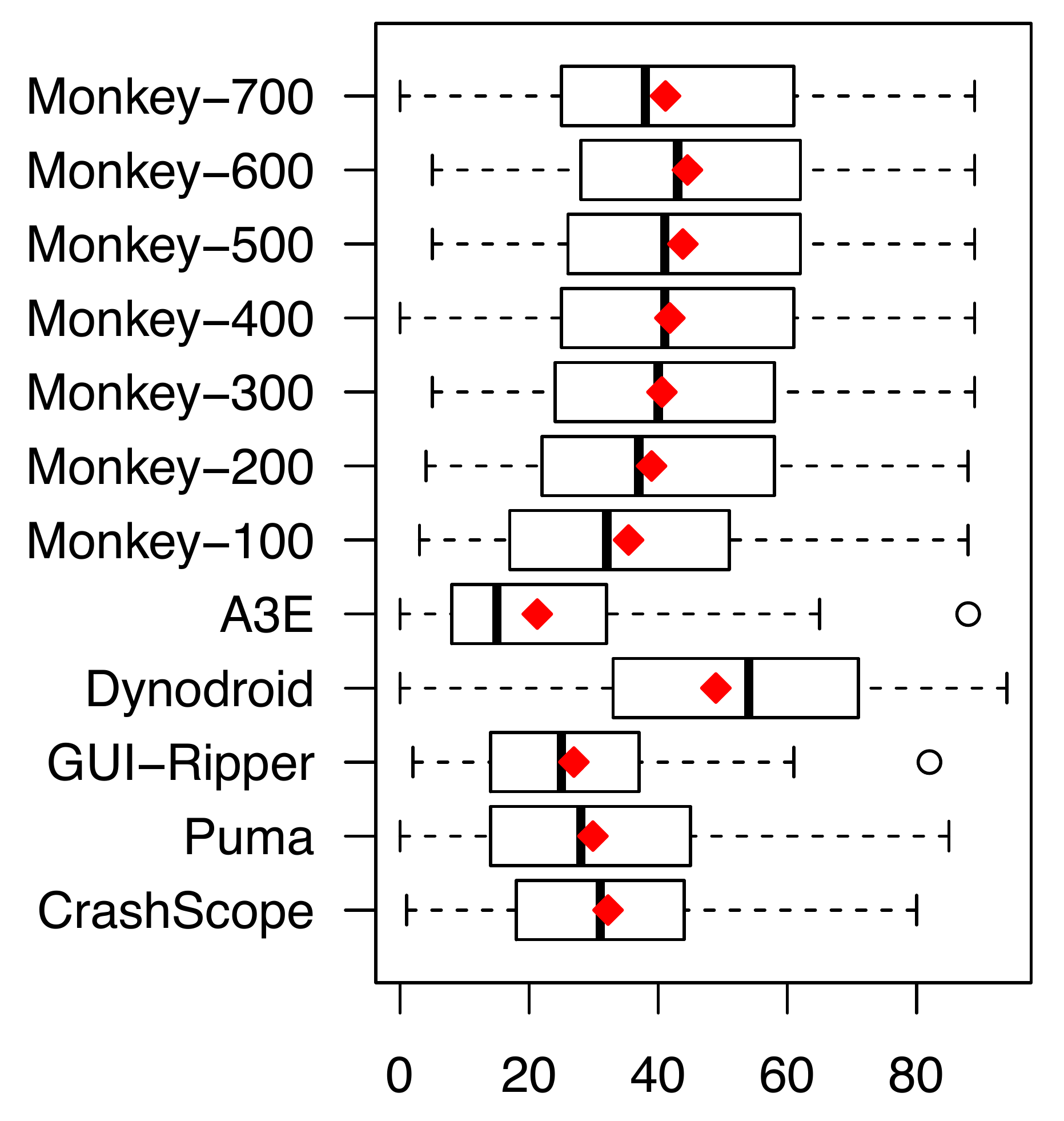}
\vspace{-0.5cm}
\caption{Average Coverage Results for the Comparative Study}
\label{fig:coverage}
\end{figure}
		
	The second observable result is that \CrashScopesp is able to detect orthogonal crashes compared to the other tools.  In order to understand why \CrashScopesp detected different crashes than the other approaches, the authors manually examined the detected crash reports to determine their causes.  Because it might not be possible to determine the exact cause or type of crashes from the other tools, we exclude a discussion here, but we speculate on the differences from \CrashScopes results. The key finding from this exploration is that \textit{the differing strategies implemented by CrashScope contributed to its ability to detect orthogonal crashes compared to the other tools.}  For instance, the crash detected by \CrashScopesp for the \textit{zooborns} app is triggered by typing unexpected text in a text box. The other tools probably missed this crash because their text generation techniques do not include unexpected inputs.  Furthermore, one aspect of this crash highlights the utility of \CrashScope's detection and reporting capabilities, namely, the thrown exception is potentially misleading to a developer.  While this crash was caused by text formatting, the exception is for an AsyncTask object, one of Android's thread handling mechanisms, meaning it could be difficult for a developer to reason about the cause of this crash in the absence of a detailed report.  Another example of an orthogonal crash discovered by \CrashScopesp is that for the \textit{PasswordMakerPro} app. While two other tools (Monkey, A$^3$E) found a crash during their exploration of this app, only CrashScope was able to discover a crash caused by a contextual feature, rotation.  This highlights the utility of the different exploration techniques.  Consequently, \textbf{RQ$_2$} can be answered as follows: \textbf{The varying strategies of CrashScope allow the tool to detect different crashes compared to those detected by other approaches.}	
	\indent The third result we see from the the crash detection data is that certain \CrashScopesp strategies are more effective at uncovering crashes than others.  The most effective of the text strategies overall was the \textit{unexpected} heuristic where all of the crashes listed for \CrashScopesp in Table \ref{tab:crashes} were uncovered, but not directly triggered by, utilizing this type of text input.
	  While the different strategies achieved similar overlapping coverage on average, different crashes were discovered during the runs of strategies where contextual features were and were not tested in adverse conditions, as discussed above, suggesting that some errors are only discoverable when contextual features are in normal states.  Overall, the \textit{forwards} heuristic for traversing the GUI led to the discovery of more crashes (8 crashes) compared to the \textit{backwards} strategy (7 crashes), with 7 of these crashes overlapping.  It should be noted that the GUI-traversal heuristic did not directly trigger any crashes directly (e.g. the changing the order of interacting with components did not lead to crashes), however, these two strategies were useful for exploring different parts of the subject applications.  
	  The most effective overall crash discovery strategy was $<$\textit{contextual\_feautres\_enabled, forward, unexpected}$>$.  Full coverage and crash results for all the tools, and all 12 of \CrashScopes strategies on a per app basis are available in our online appendix \cite{appendix}. Thus, \textbf{RQ$_3$} can be answered: \textbf{Different combinations of CrashScope strategies were more effective than others, suggesting the need for multiple testing strategies encompassed within a single tool with an emphasis on strategies for contextual features.}

The fourth observable result is that the average statement coverage of the analyzed tools (see Fig.~\ref{fig:coverage}) does not necessarily correspond to a better fault discovery capability, as \CrashScopesp was able to detect about as many crashes with lower average coverage than other tools (i.e., PUMA, Monkey, and Dynodroid).  This implies that future testing approaches for mobile apps need to take into consideration other metrics in addition to code coverage to illustrate the effectiveness of the approach. Therefore, our answer for \textbf{RQ$_4$} is: \textbf{Higher statement coverage of an automated mobile app testing tool does not necessarily imply that tool will have effective fault-discovery capabilities}.

\vspace{-0.15cm}
\section{Study 2: Reproducibility \& Readability}
\label{sec:study2}
\vspace{-0.15cm}
	The \textit{goal} of the second study is to evaluate the \textit{reproducibility} and \textit{readability} of the natural language reports generated by \CrashScopesp compared to original human written reports found in online issue trackers.  The \textit{quality focus} of this study concerns the ability of developers to reproduce bugs from \CrashScope's reports.  The \textit{context} of this study consists of eight real world Android app crashes and reports, extracted from open source apps and their corresponding issue trackers, as well as reports generated by \CrashScopesp for these same crashes (details of the crashes and corresponding apps are presented in our online appendix \cite{appendix}). In the context of this second study we examined the following RQs:

\begin{itemize}
\item \textbf{RQ$_5$}: \textit{Are reports generated with \CrashScopesp more reproducible than the original human written reports?}

\item \textbf{RQ$_6$}: \textit{Are reports generated by \CrashScopesp more readable than the original human written reports?}
\end{itemize}

\vspace{-0.1cm}
\vspace{-0.15cm}
\subsection{Methodology}
\vspace{-0.1cm}

\begin{table*}[t!]
\vspace{-0.8cm}
\centering
\caption{\textbf{User Experience Results:} This table reports the mean average response from 16 users regarding the User Experience questions posed for both \CrashScopesp generated reports and the original human written reports found in the app's issue trackers.}
\label{tab:ux}
\setlength{\tabcolsep}{0.5em}
\def\arraystretch{0.8}
\vspace{-0.3cm}
\begin{tabular}{|l|l|l|l|l|}
\hline
Question & CrashScope Mean & CrashScope StdDev & Original Mean & Original StdDev \\ \hline
\textbf{UX1}: I think I would like to have this type of bug report frequently.      & 4.00            & 0.89                          &      3.06              &          0.77                        \\ \hline
\textbf{UX2}: I found this type of bug report unnecessarily complex.                & 2.81            & 1.04                          &        2.125            &             0.96                     \\ \hline
\textbf{UX3}: I thought this type of bug report was easy to read/understand.        & 4.00           & 0.82                          &         3.00           &           0.97                       \\ \hline
\textbf{UX4}: I found this type of bug report very cumbersome to read.              &      2.50       & 1.10                          &         2.44           &            0.81                      \\ \hline
\textbf{UX5}: I thought the bug report was really useful for reproducing the crash. & 4.13            & 0.62                          &          3.44          &      0.89                            \\ \hline
\end{tabular}
\vspace{-0.6cm}
\end{table*}

	To identify the crashes used for this study, we manually inspected the issue trackers of the apps on F-droid looking for reports that described an app crash. Then, we ran \CrashScopesp on the version of the app that the crash was reported against to observe whether or not \CrashScopesp was able to capture the crash on the same emulator configuration as the previous study.  While we chose these bugs manually, the goal of this study is not to measure \CrashScope's effectiveness at discovering bugs (unlike the first study).  We acknowledge that there are situations in which \CrashScopesp will not be able to detect a fault and we outline these cases in Sec.~\ref{sec:limitations}.
	
	In order to answer \textbf{RQ$_5$} and \textbf{RQ$_6$}, we asked 16 CS graduate students from William and Mary (a proxy for developers \cite{Salman:ICSE2015}) to reproduce the eight crashes (four from the original human written reports, and four from \CrashScope).  The design matrix of this study was devised in such as way that each crash for each type of report was evaluated by four participants, no crash was evaluated twice for the same participant, and eight participants saw the human written reports first, and eight participants saw the \CrashScopesp reports first, all in the interest of reducing bias. The system names were also anonymized (\CrashScopesp to ``System A" and the human written reports to ``System B").  The full design matrix can be found in our online appendix \cite{appendix}.  During the study, participants recorded the time it took them to reproduce the crash on a Nexus 7 device for each report, with a time limit of ten minutes for reproduction.  If a participant could not reproduce the bug within the ten minute time frame or gave up in trying to reproduce the bug, that bug was marked as non-reproducible for that participant. To mitigate the ``flaky-test" problem, where outstanding factors such as Network I/O, varying sensor readings or app delay could cause difficulty of crash reproduction, when manually selecting the crashes and crash reports from the online repositories, the authors ensured that each bug was deterministically reproducible within the confines of the study environment (e.g. Using the proper version of the application that contains the bug and that the bug was always reproducible on the Nexus 7 tablet). Therefore, in order to answer \textbf{RQ$_5$}, we measured how many crashes were successfully reproduced by the participants for each type of crash report, we also measured the time it took each participant to reproduce each bug (the detailed dataset is available at \cite{appendix}).
	
	After the completion of the crash reproductions, we had each participant fill out a brief survey, answering questions regarding the \textit{user preferences} (\textit{\textbf{UP}}) and \textit{usability} (\textit{\textbf{UX}}) for each type of bug report.  We also collected information about each participants programming experience and familiarity with the Android platform.  The \textit{\textbf{UP}} questions were formulated based on the user experience honeycomb originally developed by Moville \cite{Morville:04} and were posed to participants as free form text entry questions. We forgo a discussion of the free-form question responses due to space limitations, but offer full anonymized participant responses at \cite{appendix}.  The \textit{\textbf{UX}} questions were created using statements based on the SUS usability scale by Brooke \cite{Brooke:96} and were posed to participants in the form of a 5-point Likert scale. We quantify the \textit{user experience} of \CrashScopesp and answer \textbf{RQ$_6$} by presenting the mean and standard deviation of the scores for the responses to the Likert-based questions.  The questions regarding programming experience are based on the well-accepted questionnaire developed by Feigenspan \textit{et al.} \cite{Feigenspan:ICPC12}.

\vspace{-0.15cm}
\subsection{Results \& Discussion}
\vspace{-0.1cm}

The \CrashScopesp reports achieved a similar levels of reproducibility compared to the human written reports with ~94\% (60 out of 64) of the \CrashScopesp reports being successfully reproduced by participants compared to ~92\% (59 out of 64) of the original reports.  Therefore, \textbf{RQ$_5$} can be answered as follows: \textbf{Reports generated by CrashScope are about as reproducible as human written reports extracted from open-source issue trackers}. Due to space limitations, full numerical statistics for Study 2, including time-cost to reproduce crashes, and detailed participant answers, can be accessed in our online appendix\cite{appendix}. The UX questions and results can be found in Table \ref{tab:ux}, which show that participants found \CrashScopesp reports to be more readable and useful than the original reports.  Thus, \textbf{RQ$_6$} can be answered as: \textbf{Reports generated by CrashScope are more readable and useful from a developers' perspective as compared to human written reports}. One interesting case arose from this study.  No participant assigned the original report for the C13C Schedule app was able to reproduce the bug, whereas all participants assigned the \CrashScopesp version of this app were able to reproduce it.  This is because the network needed to be disabled for the crash to manifest itself, and this was not captured in the original bug report.  This highlights the utility of \CrashScopes context-aware reports.

\vspace{-0.15cm}
\section{Limitations \& Threats to Validity}
\label{sec:limitations}
\vspace{-0.15cm}
While our empirical evaluation has shown that \CrashScopesp is effective at detecting crashes in Android apps, our tool has some inherent limitations.  \textit{First}, because \CrashScope's systematic execution engine does not implement the swipe gesture, it will not be able to execute GUI components existing within a list that does not fit entirely within the device's screen.  This limitation may cause some crashes or exceptions dependent on these types of components to be missed. The \textit{second} limitation is that \CrashScopesp does not support highly specialized text input.  This may limit the exploration capabilities of our tool for certain apps.  However, recent approaches in concolic and symbolic executions may prove useful in overcoming this limitation~\cite{Seo:FSE2014,Majumdar:ICSE2007,Jaffar:FSE2013,Yeh:SEREC2014,Mirzaei:ISSRE15}.  The \textit{third} limitation of our tool relates to window detection in Android.  Android apps are organized into screens based on activities and other windows (e.g., dialogs).  Activities are fairly simple to detect, as each has a unique name which acts as an identifier for that activity.  However, the same is not true for dialogs, as they have no unique identifier.  Each Activity can have multiple dialogs. To solve this problem we use the size of the window with the focus and in foreground as a unique identifier, as through our observations we found that very few activities employ different unique windows of the same size.  However, this is an imperfect heuristic and prone to occasional errors.  Due to checks in place in our systematic execution algorithm, this never leads to incorrect execution of the app, however, it may mean that less functionality of the app is explored compared to a method that is able to correctly identify all unique windows in an app.  

One potential threat to external validity is the fact that we used a set of 61 open source applications to evaluate \CrashScopesp in the first empirical study, and eight crashes in eight open source applications for the second empirical study.  Therefore, we can not generalize our results to Android apps in general due to the limitations of these subject apps.  However, we believe that this threat is lessened by the fact that these apps were collected from datasets in previous studies and contain several popular, complex apps.  In the context of our empirical studies, one threat to internal validity stem from the potentially surprising effects of participants in the second empirical study. To this end, there is a threat since we approximated graduate students in Computer Science as experienced Android developers.  However, this threat is mitigated by the fact that all of these participants indicated that they have extensive programming experience as well as moderate experience with the Android environment, and recent work shows that in carefully controlled experiments experienced graduate students are sufficient proxy's for developers \cite{Salman:ICSE2015}.  Another threat to internal validity concerns the manual inspection of log traces from the tools \CrashScopesp was tested against.  However, this threat is mitigated as the process was partially automated to decrease the manual examination set and the authors who examined these logs are well versed in the Android platform and automated testing approaches in research.

\vspace{-0.15cm}
\section{Conclusions}
\label{sec:concl}
\vspace{-0.15cm}
In this paper, we present \CrashScope, a practical approach for discovering, reporting, and replaying Android app crashes.  Our tool leverages a powerful algorithm for systematic exploration that is crash resilient, capable of context-aware input and text generation, and runs on a diverse set of devices and emulators.  We evaluated \CrashScopesp with respect to crash and exception detection, as compared to other state-of-the-art automatic input generation tools for Android and show that our tool is able to uncover about as many crashes as these other approaches, while offering more detailed information in the form of NL crash reports containing steps to reproduce the crash, and high-level repayable traces that can reproduce the crash on demand.  We also evaluated the \textit{reproducibility} and \textit{readability} of our automatically generated reports and show that they provide for reliable reproduction of crashes while proving more readable and usable for developers.  In the future, we aim to investigate techniques to trim bug reports, so that they contain only the necessary steps, as well as improving our systematic exploration strategy for uncovering a higher number of bugs, by adapting promising emerging approaches in model-based GUI testing \cite{Nguyen:TOSE2014}.

\vspace{-0.15cm}
\section*{Acknowledgements}
\label{sec:ack}
\vspace{-0.1cm}
This work is supported in part by the NSF CCF-1218129 and NSF CCF-1525902 grants. We would like to thank the anonymous reviewers for their insightful comments that significantly improved this paper and the authors of the Androtest benchmark tools \cite{Choudhary:ASE15} for their aid in reproducing the results.

\balance
\bibliographystyle{abbrv}
\bibliography{crashscope}

\end{document}